# Temperature-dependent exchange bias properties of polycrystalline $Pt_xCo_{1-x}$/CoO bilayers

E. Demirci[1*], M. Öztürk[1], E. Sınır[1], U. Ulucan[1], N. Akdoğan[1], O. Öztürk[1] and M. Erkovan[1,2]

[1]Gebze Institute of Technology, Department of Physics, 41400 Kocaeli, Turkey
[2]Institut für Experimentalphysik, Freie Universität Berlin, Arnimallee 14, 14195 Berlin, Germany.

## Abstract

Temperature and PtCo composition dependence of exchange bias in $Pt_xCo_{1-x}$/CoO bilayers is investigated. It is observed that exchange bias properties, blocking temperature and magnetic anisotropy of the $Pt_xCo_{1-x}$/CoO thin films are strongly affected by the concentration of Pt at the common interface. A detailed structural analysis of antiferromagnetic CoO has been done by x-ray photoelectron spectroscopy (XPS). The XPS data revealed that CoO layer grows with non-stoichiometric behavior which results in other cobalt oxide phases and thus lowers blocking temperature for the exchange biased systems. Increase of Pt concentration in PtCo layer strengths the exchange bias and thus decreases training of the system. It also results in enhancement of growth-induced uniaxial magnetic anisotropy and onset of exchange bias at higher temperatures. The role of Pt concentration on the interfacial interactions between ferromagnetic and antiferromagnetic layers, and the effect of superstoichiometric CoO on blocking temperature are also discussed.

# 1. Introduction

In recent years, magnetic read heads [1, 2], magnetic random access memories (MRAM) [1-3] and magneto-electronic switching devices (spin valves) [2, 4-7] have received much attention due to their applications in data storage technology. One of the key elements in development of such devices is exchange bias (EB) effect. Exchange bias, also known as unidirectional magnetic anisotropy, is a resultant effect of interfacial exchange coupling between ferromagnetic (FM) and antiferromagnetic (AF) layers after cooling the system below to the Néel temperature under an external magnetic field [8, 9]. Due to this unidirectional anisotropy, FM hysteresis loop is generally shifted to the negative direction of the cooling field in the magnetic field axis.

There are many studies related to exchange bias in the literature with different systems. Some of them are Co/CoO [8-11], Fe/CoO [12-14], Ni/NiO [15, 16], Fe/FeO [17], Fe/Fe$_3$O$_4$ [18], Fe/NiMn [19], Ni–Mn–Sb [20] and Ni–Mn–Sn [21] Heusler alloys as well as Sm$_{0.5}$Ca$_{0.5}$MnO$_3$ [22] manganites nano particles. As a ferromagnetic material Fe [4, 13, 23-26], Co [2, 10, 11, 27, 28] and their alloys are extensively studied to enhance the understanding of this phenomenon. In addition, different AF materials such as IrMn [29], FeO [30], NiO [31], Fe$_2$O$_3$ [32], and CoO [8-11, 13, 14] are used. Although CoO is not used as an AF material in real devices, due to its strong anisotropy and convenient Néel temperature ($T_N$= 291 K), which is very close to the room temperature, CoO becomes very preferable among other AF materials for laboratory measurements. In addition to the FM and AF material type, exchange bias field and blocking temperature ($T_B$) of an exchange bias system are also strongly affected by both FM and AF layer thicknesses [33-35], the stoichiometry of the AF material [35] and the number of exchange biased interfaces in multilayered systems [11].

In this study, we have investigated the exchange bias properties of polycrystalline $Pt_xCo_{1-x}$/CoO bilayers grown by using magnetron sputtering technique. FM and AF layer thicknesses are fixed as 15 nm for both samples and composition of FM PtCo layer is changed. The chemical stoichiometry and the atomic concentrations within individual layers and the thicknesses of the samples are determined by using both x-ray photoelectron spectroscopy and transmission electron microscopy. Room and low temperature magnetization behaviors, exchange bias properties and training effects of the samples were investigated by vibrating sample magnetometry and ferromagnetic resonance. We also discussed possible reasons for the enhanced exchange bias and blocking temperature as well as decreased training by increase of Pt concentration in FM layer.

## 2. Sample Preparation and Structural Characterization

The samples were fabricated and characterized in a cluster chamber combined with magnetron sputtering deposition chamber and analytical chamber. Both are connected to a load-lock chamber. Multilayer film samples were grown onto naturally oxidized p-type Si (001) substrates by magnetron sputtering with base pressure $<1\times10^{-8}$ mbar. All substrates were subjected to cleaning process such as ethanol and methanol bath, and were transferred into UHV conditions for annealing up to 600 °C. They were hold at this temperature for 30 minutes as a final cleaning step. For magnetron sputtering depositions, Ar process gas (of 6N purity) was exposed to deposition chamber so that the base pressure level became $1.2\times10^{-3}$ - $1.3\times10^{-3}$ mbar during growth. Three-inch cobalt (3N5) and platinum (4N) elemental targets were used to grow the layers of PtCo alloys and CoO. The distance between the target and substrate was always kept in 100 mm for all growth processes. The ratio of oxygen (of 6N purity) floating with Ar was controlled by mass flow controller (MKS 1179A) to optimize CoO layer. The PBN heater, at the sample holder, located under the substrates has the capability of annealing up to 1000 °C in range from UHV condition to reactive condition ($10^{-5}$ mbar with $O_2$). The sample holder was cooled by chilled water to stabilize the sample temperature during annealing process. During deposition, QCM (quartz crystal monitoring) thickness monitor was used to observe deposition ratio in situ. Thickness calibration for QCM was done by using XPS (X-ray photoelectron spectroscopy), so both thicknesses measured by QCM and calculated from photoemission attenuations are the same. Pt and Co film deposition ratios were calibrated before synthesizing PtCo/CoO bilayers. The parameters (such as tooling factor and new impedance values) of QCM are calibrated by photoemission attenuations besides its default values. In order to determine the deposition ratios of the Pt and Co films, the pure silver substrate is used to observe the attenuations of silver atoms photoemission as a function of additional material onto the substrate. For thickness determination using XPS signals both Ag $3d_{5/2}$ attenuation as a function of Pt and Co exposure, as well as the Pt $4f_{7/2}$ and Co $2p_{3/2}$ to Ag $3d_{5/2}$ intensity ratio for a given films were compared. In converting these ratios to films thickness, the electron mean free path was calculated by using the TPP formula [36]. The multilayer composition is Cr/PtCo/CoO/Si(001). The top layer is 15 nm Cr grown from pure Cr target to prevent oxidation during transferring of the samples between the growth chamber and the magnetic characterization system. The thicknesses of PtCo and CoO layers are both 15 nm.

PtCo alloy films were grown using repeated deposition sequences of sub-monolayers of Pt and Co, since the circular magnetron guns faces down to deposition surfaces in right angle. The number of sequences was changed based on the thickness of films. The shutters in front of each gun were opened during each sequence of depositions when the sample on the holder was moved under certain gun of Pt or Co. The period of shutter in open position was calculated with respect to deposition ratio of gun loaded with Pt or Co. The power applied for Pt target (DC gun) and Co target (RF gun) were 2 watt and 25 watt, respectively. They cause the deposition rates of 0.1 Å/sec for Pt 0.3 Å/sec for Co. Based on deposition rates for every elemental deposition, the number of atoms of Pt and Co for each alloy type can be calculated per unit area per time. According to the elemental stoichiometry, the deposition periods of Pt and Co targets for each sequence are tabulated in Table 1.

**Table 1.** Deposition periods for Pt and Co targets.

| Alloy | Pt deposition time (sec) | Co deposition time (sec) |
|---|---|---|
| $Pt_{40}Co_{60}$ | 4 | 4 |
| $Pt_{55}Co_{45}$ | 5 | 3 |

Therefore, the amount of film thickness for every sequence becomes less than half of a monolayer. During sequential depositions, both Pt and Co targets were operated at the same time and the temperature of the substrate was 350°C. After the growth process, chemical stoichiometry was characterized by using XPS in situ. The selected stoichiometries of alloy layers for respective samples were $Pt_{0.40}Co_{0.60}$ and $Pt_{0.55}Co_{0.45}$.

Fig. 1 shows survey XPS spectra from alloy surfaces of respective samples. The XPS spectra were taken after the removal of about 15 nm Cr cap layer by Ar sputtering. The cap layer removal process took place in analytical chamber. High purity Ar (6N) gas was leaked through a precision leak valve into the analytical chamber so that the pressure was stabilized at $1 \times 10^{-5}$ mbar during sputtering of the surface. The base pressure of chamber was always $<5 \times 10^{-10}$ mbar. Besides the survey XPS spectra, high resolution window XPS spectra for the major photoemission Pt 4f and Co 2p regions were also taken to calculate Pt-Co ratios in the PtCo alloy layers. The integrated areas of Pt 4f and Co 2p peaks were calculated by CASA XPS 2.3.14 commercial software (SPEC GmbH). The Shirley background function was used for fit analyses of peaks. The Voigt function corresponding to photoemission nature was used for calculating the peak area. Peak areas of Pt 4f and Co 2p were divided by tabulated atomic sensitivity factors since every element has a different sensitivity within photoemission process. The calculated Pt to Co ratios within PtCo alloy layers of the two samples are 40:60 and 55:45, respectively.

The CoO layer was studied more carefully due to the possible existence of other oxide forms of Co. Since oxidation process of Co results in different thicknesses, the calculated CoO film thickness (from attenuation of photoelectrons in conjunction with QCM) was confirmed by Veeco Profilometer. According to the profilometer study, the thickness of 15 nm was succeeded with 300 sec deposition.

In order to grow the CoO film layer, the pure oxygen molecular gas (6N Grade) was released (by mass flow meter fixed at 0.15 sccm) into the growth chamber so that the chamber pressure was established at $5 \times 10^{-5}$ mbar. After stabilization of partial oxygen pressure, Ar process gas (by mass flow meter fixed at 2.6 sscm) was released into the chamber so that the total vacuum level was fixed in the range of $1.2 \times 10^{-3}$-$1.3 \times 10^{-3}$ mbar. At this process pressure, the rf-sputter gun loaded by elemental Co target was fired with 40 watt. The shutter in front of the source was opened 10 sec after the source was fired.

Fig. 2 shows the XPS data of CoO surface to characterize oxidations. The satellite peaks shown in Fig. 2(a) most likely indicate successful $Co^{2+}$ oxidation. The satellite peaks associated with Co 2p peaks appeared stronger when the oxide form became CoO rather than the other forms of Co-oxide because of the charge-transfer band structure characteristic of the late 3d transition metal oxide [37]. Due to two oxide formations, Co 2p peak is broader compared to the peak of a single oxide surface. The width of the Co 2p photoemission peaks is consistent with the presence of both $Co^{2+}$ and $Co^{3+}$ as well as the $Co^{2+}$ satellites. The one sits on 779.3 eV binding energy, and the other one sits on 781.7 eV binding energy as seen in Fig. 2. Both correspond to $Co^{2+}$ and $Co^{3+}$ respectively. In the region of both Co $2p_{3/2}$ and

Co $2p_{1/2}$ peak lines, two Voigt peaks were fitted to calculate ratio of Co-oxide formation. The result shows that $Co^{2+}$ / $Co^{3+}$ ratio is 2.6. This means that Co-oxide layer contains mostly CoO (78%) formation. The O 1s XPS spectrum (Fig. 2(b)) shows a main peak at 529.5 eV. This main peak indicates main CoO formation. This O 1s at 529.5 eV is also observed for $Co_3O_4$ [38]. However, there is another small peak sitting at 531.4 eV as seen in Fig. 2. It would be easy to attribute this to chemisorbed $Co^{3+}$, $OH^{-1}$ or $O^{-2}$ which were observed in the previous works [39, 40]. Carson's HREELS work and Tyuliev's angle resolved XPS works [41] indicated that O 1s at 531.4 eV was related to near surface region and was not from surface contaminations. It was nature of Co-oxide surface.

In order to be sure about the thicknesses of the samples and to compare them with XPS results, we have taken cross-sectional images of the samples by using transmission electron microscopy (TEM). The TEM picture of the Cr/PtCo/CoO/Si(001) sample stack is shown in Fig. 3. The Pt top layer is deposited prior to FIB (focused ion beam) preparation of the sample cross section in order to prevent charging effects. TEM data show that AF CoO, FM $Pt_{0.40}Co_{0.60}$ and Cr layers have thicknesses of 14.8 nm, 14.7 nm and 16.9 nm, respectively. Since the TEM picture has low resolution, detailed information about interface quality between FM and AF layers cannot be provided.

## 3. Magnetic Anisotropies

In order to determine the effect of Pt concentration on magnetic anisotropies, the samples have been investigated by using ferromagnetic resonance (FMR) technique at room temperature (RT). FMR measurements were carried out by using Bruker EMX electron spin resonance (ESR) spectrometer operating at X-band (9.8 GHz). An electromagnet with a magnetic field up to 2.2 Tesla provides the dc magnetic field to this spectrometer. Magnetic field component of the X-band microwave field is perpendicular to the dc magnetic field. A goniometer was used to rotate the sample holder in the static magnetic field which is parallel to the film plane. This is called as in-plane geometry and for this geometry the microwave magnetic field is perpendicular to the film plane whereas the static magnetic field is in the film plane. FMR signal intensity is recorded as a function of the static magnetic field at a given in-plane magnetic field angle ($\varphi$).

FMR resonance fields of $Pt_{0.40}Co_{0.60}$/CoO and $Pt_{0.55}Co_{0.45}$/CoO samples obtained from static magnetic field sweeps have been plotted as a function of in-plane magnetic field angle ($\varphi$), as seen Fig. 4. It is obvious from the data that 0° and 180° directions correspond to minimum resonance fields forming an easy-axis whereas 90° and 270° with maximum resonance fields form a hard-axis of magnetization leading to a uniaxial in-plane magnetic anisotropy. Both samples have nearly the same minimum resonance field values. However, maximum resonance field values are different for the samples. Less Pt containing sample ($Pt_{0.40}Co_{0.60}$/CoO) has a lower resonance value as 980 Oe and higher Pt containing sample ($Pt_{0.55}Co_{0.45}$/CoO) has higher value as 1050 Oe. The samples have polycrystalline structure and observation of uniaxial magneto crystalline anisotropy is not expected. This unexpected behavior of PtCo/CoO polycrystalline thin films is explained by growth conditions. This type of anisotropy is called as growth induced or geometric (oblique) anisotropy [42, 43] and previously reported for polycrystalline [Co/CoO]$_n$ multilayered thin films [11]. It is also important to note that FMR measurements have been carried out at room temperature. Thus, the difference in the magnetic anisotropy of the samples can only be attributed to PtCo composition, but not to the exchange bias. Indeed, enhancement of magnetic anisotropy by increasing Pt concentration is not surprising and it is also observed in ultra-thin Co/Pt films [44].

After determining magnetic anisotropies, we have studied exchange bias properties of two different $Pt_xCo_{x-1}$/CoO samples by using Quantum Design PPMS 9T vibrating sample magnetometer (VSM) at the easy axis. Since the Néel temperature is about 290 K for antiferromagnetic CoO layer, the samples were heated up to 320K before cooling down to a target measurement temperature. This heating and recooling procedure was repeated to perform magnetization measurements at each target temperature (10 to 305 K) to eliminate the training effect. An in-plane magnetic field of 2 kOe was applied while cooling the samples to the target temperatures. Then, magnetization versus magnetic field hysteresis loops were taken at this defined temperatures as shown in Fig. 5. Hysteresis loops at 10 K show negative exchange bias with exchange bias fields of magnitudes 140 Oe and 232 Oe for $Pt_{0.40}Co_{0.60}$/ CoO and $Pt_{0.55}Co_{0.45}$/ CoO, respectively. These values have been calculated according to well-known formula of exchange bias,

$$H_{EB} = \frac{H_{C1} + H_{C2}}{2} \qquad (1)$$

where $H_{C1}$ and $H_{C2}$ are the coercive field values of the shifted hysteresis loop.

Fig. 5 presents that at 305 K the hysteresis loops are symmetric. After field cooling, however, the hysteresis loops become asymmetric at 10 K and shift to the negative magnetic field direction. After calculating the exchange bias field ($H_{EB}$) values from the coercive fields of each hysteresis loop, we plotted temperature dependence of exchange bias and coercive fields in Fig. 6. Fig. 6 shows that the exchange bias vanishes above a temperature often called as blocking temperature ($T_B$) and for both samples it is lower than the Néel temperature of bulk antiferromagnetic CoO. Generally thick AF layers have equal blocking and Néel temperatures ($T_B \approx T_N$) [45-47] and very thin AF layers have lower blocking temperatures ($T_B < T_N$) [48, 49]. In this study, from RT to the $T_B$ magnitudes of the coercive fields increase slowly with decreasing temperature, but they are equal to the each other. Well below the blocking temperature $H_{C1}$ becomes larger in magnitude than $H_{C2}$ with a negative exchange bias in this region (Figs. 6a and 6b). $T_B$ values are around 200 K and 250 K for $Pt_{0.40}Co_{0.60}$/CoO and $Pt_{0.55}Co_{0.45}$/CoO, respectively. This suggests that by increasing Pt concentration from 40 to 55% in PtCo layer, $T_B$ increases by an amount around 50 K. Fig. 6c shows the drastic change in the slope of exchange bias below the blocking temperature for both samples.

The exchange bias and blocking temperature dependences on Pt concentration for both $Pt_xCo_{x-1}$/CoO samples are compared. With increasing Pt concentration, the amount of Co atoms for the same total thickness decreased and as a consequence the rate of Co magnetic moments decreased in the total magnetization. On the other hand, the increase of Pt concentration for our systems caused the increase of exchange bias field and blocking temperature. Table 2 shows the values of $-H_{EB}$ and $T_B$ determined from the hysteresis loops for two different Pt concentrations. Obviously these values were increased by adding more Pt to $Pt_xCo_{1-x}$/CoO system.

**Table 2**. Dependence of exchange bias field and blocking temperature on Pt concentration for $Pt_{0.4}Co_{0.6}$/CoO and $Pt_{0.5}Co_{0.5}$/CoO samples.

| Pt Concentration | $-H_{EB}$ (Oe) at 10 | $T_B$ (K) |
|---|---|---|
| 0.40 | 140 | 200 |
| 0.55 | 232 | 250 |

## 4. Training Effect Measurements

It is known that magnitude of the exchange bias field, $-H_{EB}$, and $H_C$ (half-width of the hysteresis loop) often decrease monotonically due to consecutive hysteresis loops with increasing cycling number (n) [50, 51]. This effect is generally called as training effect. Since exchange bias has important technological applications, a clear understanding of training effect could lead to technological advances. In order to investigate the training effect of $Pt_xCo_{x-1}$/CoO system, the samples were cooled down from 320 K (above $T_N$) to target 10 K under a magnetic field of 2 kOe and consecutive hysteresis loops were taken at this temperature for both samples. The number of cycled loops is denoted as n, and it is about 40 for our samples. Fig. 7 shows the dramatic change of the successive hysteresis loops measured at 10 K. The first hysteresis loop has a huge asymmetry, but after consecutive hysteresis loops $-H_{C1}$ gradually decreases for increasing n. As seen in Fig. 7 for the first cycle $-H_{C1}$ (n=1) is about 638 Oe and it decreases to 588 Oe for the second cycle (n=2). Large change in coercive fields is only observed between first two measurements (n = 1 and n = 2), which is expected and observed in a previous study [52]. For n≥2, $-H_{EB}$ decreases gradually with increasing number of cycled loops.

Fig. 8 shows the gradual decrease in the exchange bias field after about 40 consecutive hysteresis loops for both samples at 10 K. However, the important thing to take into consideration is that exchange bias field of the sample with higher Pt concentration decreases slowly compared to the other one. Its maximum and minimum exchange bias field values are about 228 Oe and 198 Oe, respectively. The difference between maximum and minimum values ($\Delta H_{EB}$) is about 30 Oe. On the other hand, maximum and minimum exchange bias field values for the other sample are about 138 Oe and 80 Oe, respectively. The difference between maximum and minimum values ($\Delta H_{EB}$) is about 58 Oe. This data shows that increase in Pt concentration, decreases training effect and results in stronger exchange bias.

## 5. Discussion

The magnetization measurements showed that the $T_B$ of both samples is lower than the expected bulk value. As discussed before, this value is usually very close to the Néel temperature of AF material. Two reasons can be considered for this low $T_B$ of PtCo/CoO samples. The first one is that the blocking temperatures are strongly related with the AF layer thickness, especially for very thin films. According to the literature [8, 23, 24], the lower limit for CoO is 20 nm to behave like bulk of itself. $T_B$ reduces with decreasing CoO thickness below this limit. The second reason is the superstoichiometric structure of CoO. It is reported that only stoichiometric CoO can exhibit the highest $T_B$ close to the $T_N$ of bulk [10, 13, 24, 26, 35, 53]. Since we observed secondary oxide phases in CoO layer, this also could lead to low $T_B$ [11]. We also observed that increasing Pt concentration within the FM PtCo layer elevates the blocking temperature to higher values, extending the temperature interval for the observation of the exchange bias. To the opposite direction, as the Pt concentration is lowered, blocking temperature is getting reduced. This, at first sight, might lead to the idea that the presence of Pt in FM layer which was otherwise a pure Co layer, is also a factor in reduced blocking temperature in these $Pt_xCo_{1-x}$/CoO systems in addition to AF layer properties discussed above. However, if we extrapolate this reduced blocking temperature behavior to successively lower amounts of Pt leading eventually to almost 100% Co FM layer, we will reach the bottom line and our FM/AF system will already possess the lowest blocking temperature. Therefore, we can conclude that the existence of Pt in PtCo layer is not likely to be responsible for the reduced blocking temperature. Its presence, as our measurements with our two samples having different amounts of Pt suggest, can only lead to higher values of the blocking temperature fixed on the upper limit by the Néel temperature $T_N$. Yet, one should be cautious in extrapolating the results gathered from two samples with a minimum Pt concentration of 40% in PtCo layer. Further studies in these systems with lower

Pt concentrations are required  We also speculate that the enhanced magnetic anisotropy with increasing Pt concentration in PtCo layer, as revealed by in-plane FMR data, can explain why $T_B$ is elevated as Pt to Co ratio gets higher. Early in this paper, we presented that minimum resonance field values of the two samples at their own easy-axes were nearly the same while the sample with higher amount of Pt had a higher resonance magnetic field value along the in-plane hard-axis of magnetization. This was an enhanced magnetic anisotropy for increasing Pt concentrations. That said, one can argue that the increased resonance fields were along the hard-axis while the different $T_B$ values were extracted from easy-axis hysteresis measurements of respective samples along which resonance values were the same for both samples. How could we, then, explain different $T_B$ values of the samples in the light of FMR measurements? A possible explanation might be that as the external magnetic field direction in hysteresis measurement is reversed, the macroscopic FM layer magnetization will try to rotate in plane to the new direction and in the course has to pay a visit to the in-plane hard-axis. The sample with higher Pt concentration has a higher hard-axis anisotropy that would make the passing of the magnetization more difficult. It is not surprising to think, then, that the increased anisotropy could lead to a stronger tendency for magnetization in FM layer to stay close to the unidirectional anisotropy direction that we try to induce by field cooling procedure. As more Pt contributes to the interfacial magnetic coupling, there is a stronger anisotropy and it becomes harder for thermal agitations to randomize spin alignments away from preferred direction of the cooling field and to block the onset of exchange bias, and thus the EB properties can arise at even higher temperatures. Of course, the reason for the enhanced in-plane hard-axis anisotropy for increasing Pt to Co ratios in $Pt_xCo_{1-x}$/CoO bilayers and how this can translate to elevated blocking temperatures and higher exchange bias field values require a more microscopic treatment than this model has to offer.

We have also observed that exchange bias field value depends on the ratio of platinum. In order to understand this behavior we can at least discuss the following scenario. It is generally assumed that the origin of exchange bias is an interfacial effect between ferromagnetic and antiferromagnetic layers. The interface effect is microscopically related to exchange energy which defined by following equation,

$$E_{ex} = -2J_{ex}\,S_i S_j \cos\phi \qquad (2)$$

where $J_{ex}$ is a particular integral, called as the exchange integral and $\phi$ is the angle between spins. For the exchange bias systems, $S_i$ and $S_j$ can be regarded as magnetic moments of ferromagnetic and antiferromagnetic layers, respectively. Specifically for PtCo/CoO system, $S_j$ can be attributed to Co atoms at the antiferromagnetic CoO layer. On the other hand, $S_i$ cannot be attributed only to Co atoms at the ferromagnetic PtCo layer, but also to Pt atoms which are acquired to have magnetic characteristic. Hence, at the interface of FM/AF system, Co magnetic moments at the antiferromagnetic layer should interact with Pt magnetic moments as well as Co magnetic moments at the ferromagnetic layer. According to experimental results, the strength of the exchange coupling between FM/AF layers increases with increasing of Pt concentration. This indicates that Pt atoms make a major contribution to the exchange energy at the interface. For this reason we observed higher exchange bias field for $Pt_{0.55}Co_{0.45}$/CoO bilayer.

## 6. Conclusions

Exchange bias properties of polycrystalline $Pt_xCo_{1-x}$/CoO bilayers are studied as a function of temperature and Pt concentration. XPS data show that presence of secondary oxide phases within the CoO layer reduces blocking temperature of the exchange biased bilayers. FMR experiments present uniaxial in-plane magnetic anisotropy at room temperature. This behavior becomes stronger when the Pt concentration is increased. Moreover, temperature-dependent magnetization measurements demonstrate that strength and onset temperature of exchange bias are enhanced by increasing Pt concentration. Thus, the training of exchange bias is reduced. These results show that manipulation of common interface between ferromagnetic and antiferromagnetic layers, gives possibility to tune exchange bias.

## Acknowledgements

The authors wish to acknowledge Ömer Faruk Deniz for taking TEM image of samples, and Melek Türksoy Öcal for her help to perform XPS measurements. This work was supported by TÜBİTAK (The Scientific and Technological Research Council of Turkey) through the project numbers 112T857, 106T576 and 212T217.

## List of Figure and Table Captions

**Fig. 1.** XPS survey spectra from the alloy surfaces of the two samples having different Pt to Co ratios. The ratios of peak areas under the Pt 4f and Co 2p regions provide the ratio of Pt and Co atoms. The one shown by dark blue line has 55% concentration of Pt and 45% concentration of Co in Pt-Co alloy, and the other one (light red color line) has 40% concentration of Pt and 60% concentration of Co in Pt-Co alloy.

**Fig. 2.** XPS survey spectra from the Co-oxide surface and the Co 2p spectral region (Fig. 2a) showing the effects of oxidation. The observed satellites (noted with stars) are consistent with the presence of CoO formations in oxide layers of the film. O 1s region spectra (Fig. 2b) is showing a small shoulder noted with a star. Full XPS survey spectra from oxide layer is shown in Fig. 2c.

**Fig. 3.** TEM image of Cr/PtCo/CoO/Si(001) sample stack for $Pt_{0.40}Co_{0.60}$ concentrations.

**Fig. 4.** The FMR resonance fields of $Pt_{0.40}Co_{0.60}$/CoO and $Pt_{0.55}Co_{0.45}$/CoO samples as a function of in-plane magnetic field angle ($\varphi$).

**Fig. 5.** Symmetric hysteresis loops at 300 K and asymmetric hysteresis loops at 10 K for $Pt_{0.40}Co_{0.60}$/CoO and $Pt_{0.55}Co_{0.45}$/CoO. (Points are the data and lines are guide to the eyes.) a) For $Pt_{0.40}Co_{0.60}$/CoO, coercive field values at RT are $H_{C1}$= -155 Oe, $H_{C2}$= 155 Oe with no EB field whereas they become $H_{C1}$= -609 Oe, $H_{C2}$= 329 Oe with $H_{EB}$= -140 Oe at 10 K. b) For $Pt_{0.55}Co_{0.45}$/CoO, coercive field values are $H_{C1}$= -66 Oe, $H_{C2}$= 66 Oe with no EB field at RT and they are $H_{C1}$= -603 Oe, $H_{C2}$= 139 Oe with $H_{EB}$= -232 Oe at 10 K.

**Fig. 6.** The change of coercive and exchange bias fields as a function of temperature from 10 K to 320 K (Points are the data and lines are guide to the eyes.). a) Temperature dependence of coercive fields for $Pt_{0.40}Co_{0.60}$/CoO. $T_B$ is marked around 200 K from the splitting of -$H_{C1}$ and $H_{C2}$ fields. b) Temperature dependence of coercive fields for $Pt_{0.55}Co_{0.45}$/CoO. $T_B$ is marked around 250 K from the splitting -$H_{C1}$ and $H_{C2}$ fields. c) –$H_{EB}$ fields increase dramatically below $T_B$ for both samples.

**Fig. 7.** Training effect measurements of $Pt_{0.40}Co_{0.60}$/CoO sample taken at 10K. The number of consecutive hysteresis loops is 40. The main graph shows the drastic change in $H_{c1}$ and inset shows the full hysteresis loops.

**Fig. 8.** Training of exchange bias ($H_{EB}$) as a function of cycle number (n) for $Pt_{0.40}Co_{0.60}$/CoO and $Pt_{0.55}Co_{0.45}$/CoO samples at 10 K.

**Table 1.** Deposition periods for Pt and Co targets.

**Table 2**. Dependence of exchange bias field and blocking temperature on Pt concentration for $Pt_{0.4}Co_{0.6}$/CoO and $Pt_{0.5}Co_{0.5}$/CoO samples.

**Figures and Tables**

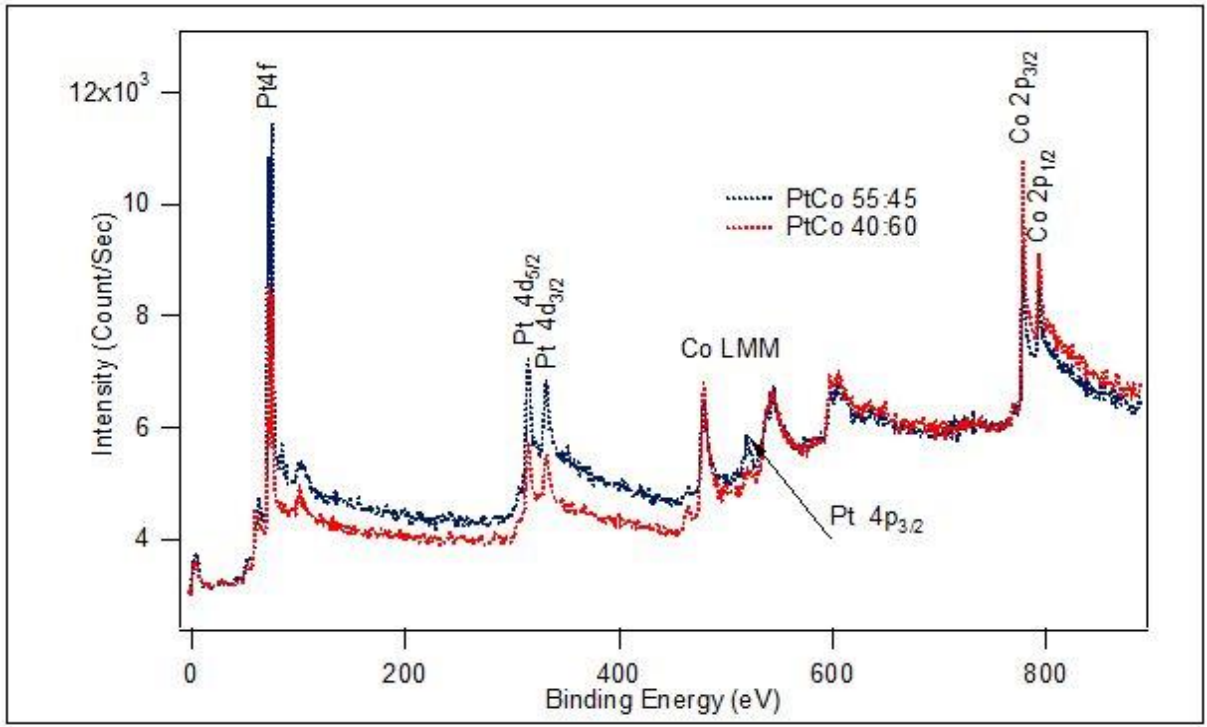

**Fig. 1.** XPS survey spectra from the alloy surfaces of the two samples having different Pt to Co ratios. The ratios of peak areas under the Pt 4f and Co 2p regions provide the ratio of Pt and Co atoms. The one shown by dark blue line has 55% concentration of Pt and 45% concentration of Co in Pt-Co alloy, and the other one (light red color line) has 40% concentration of Pt and 60% concentration of Co in Pt-Co alloy.

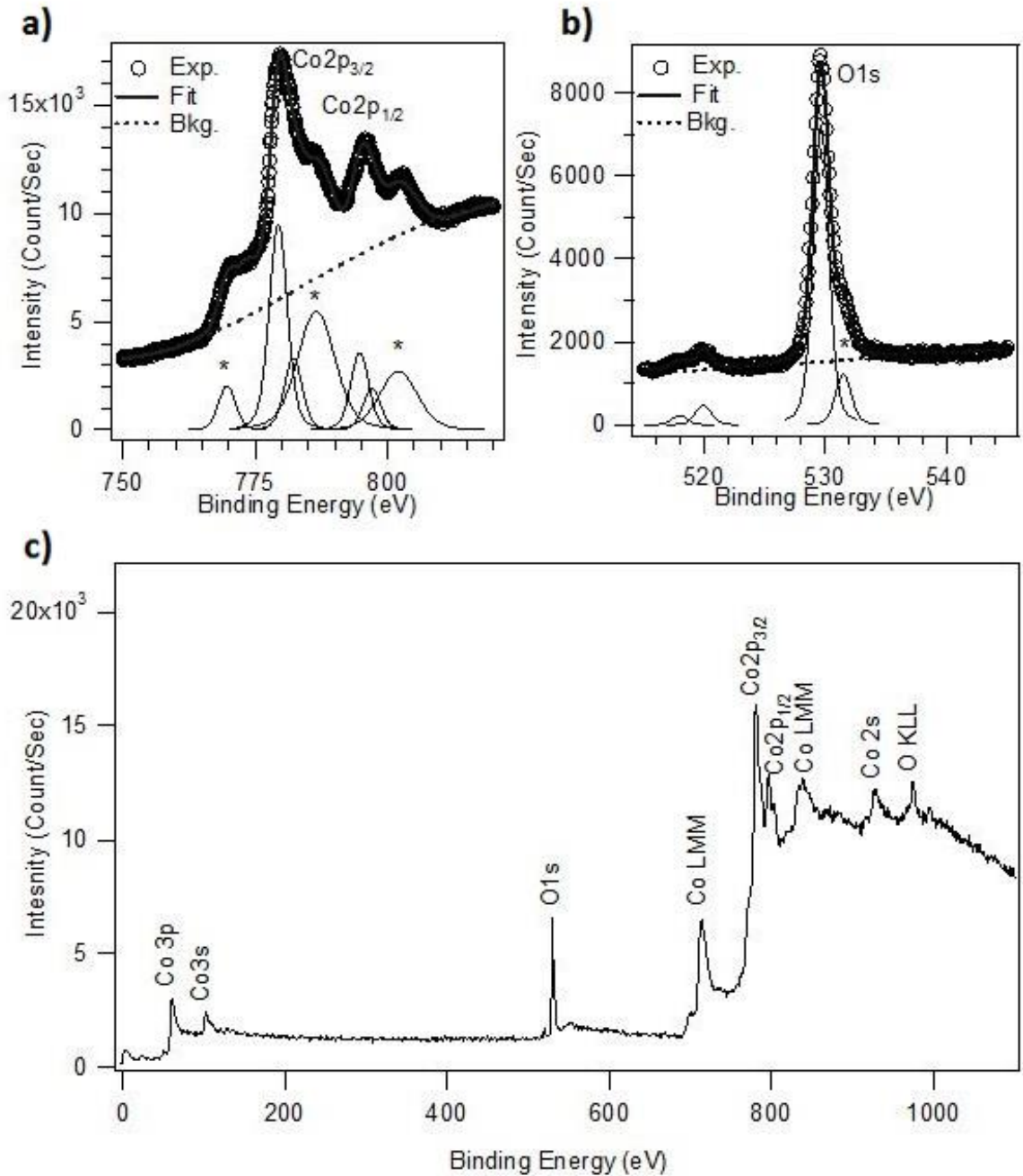

**Fig. 2.** XPS survey spectra from the Co-oxide surface and the Co 2p spectral region (Fig. 2a) showing the effects of oxidation. The observed satellites (noted with stars) are consistent with the presence of CoO formations in oxide layers of the film. O 1s region spectra (Fig. 2b) is showing a small shoulder noted with a star. Full XPS survey spectra from oxide layer is shown in Fig. 2c.

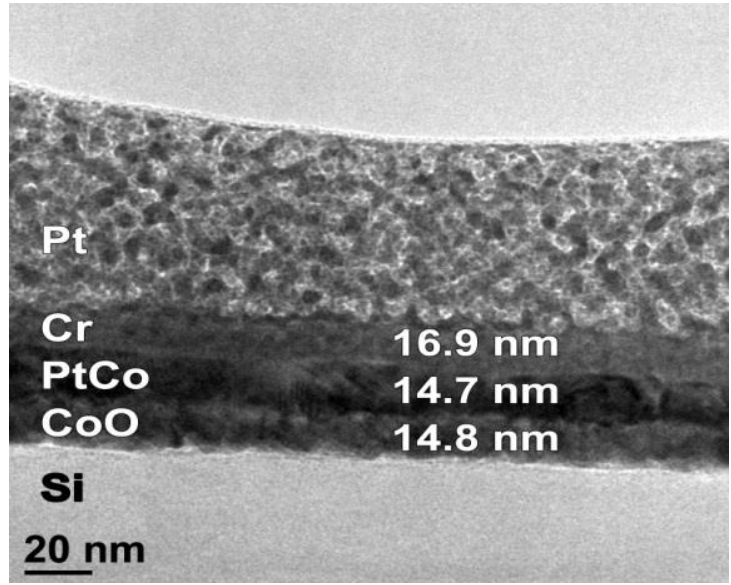

**Figure 3.** TEM image of Cr/PtCo/CoO/Si(001) sample stack for $Pt_{0.40}Co_{0.60}$ concentrations.

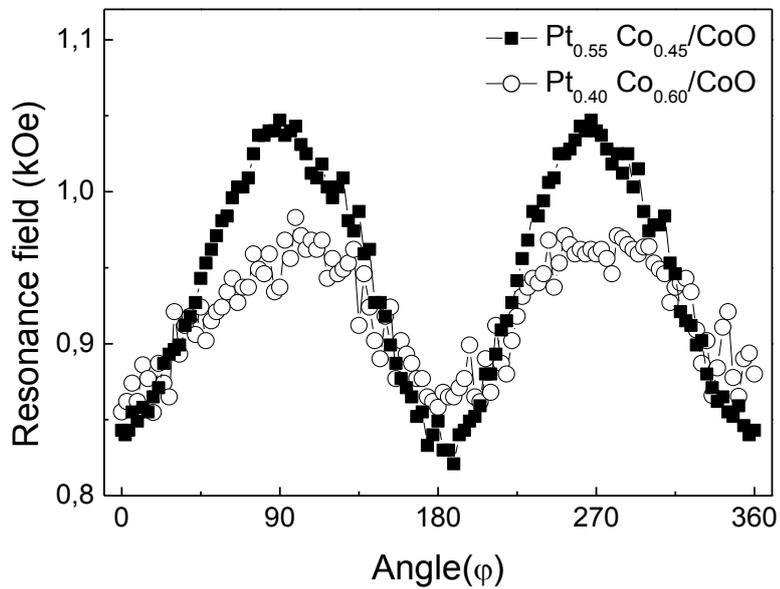

**Fig. 4.** The FMR resonance fields of $Pt_{0.40}Co_{0.60}$/CoO and $Pt_{0.55}Co_{0.45}$/CoO samples as a function of in-plane magnetic field angle ($\varphi$).

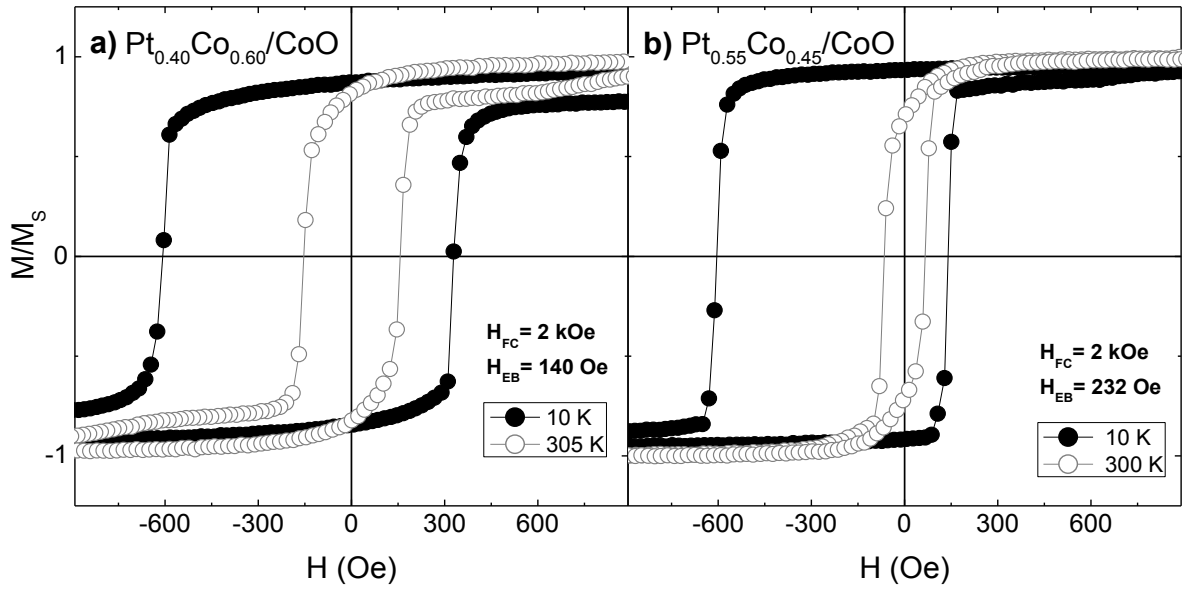

**Fig. 5.** Symmetric hysteresis loops at 300 K and asymmetric hysteresis loops at 10 K for $Pt_{0.40}Co_{0.60}/CoO$ and $Pt_{0.55}Co_{0.45}/CoO$. (Points are the data and lines are guide to the eyes.) a) For $Pt_{0.40}Co_{0.60}/CoO$, coercive field values at RT are $H_{C1}$= -155 Oe, $H_{C2}$= 155 Oe with no EB field whereas they become $H_{C1}$= -609 Oe, $H_{C2}$= 329 Oe with $H_{EB}$= -140 Oe at 10 K. b) For $Pt_{0.55}Co_{0.45}/CoO$, coercive field values are $H_{C1}$= -66 Oe, $H_{C2}$= 66 Oe with no EB field at RT and they are $H_{C1}$= -603 Oe, $H_{C2}$= 139 Oe with $H_{EB}$= -232 Oe at 10 K.

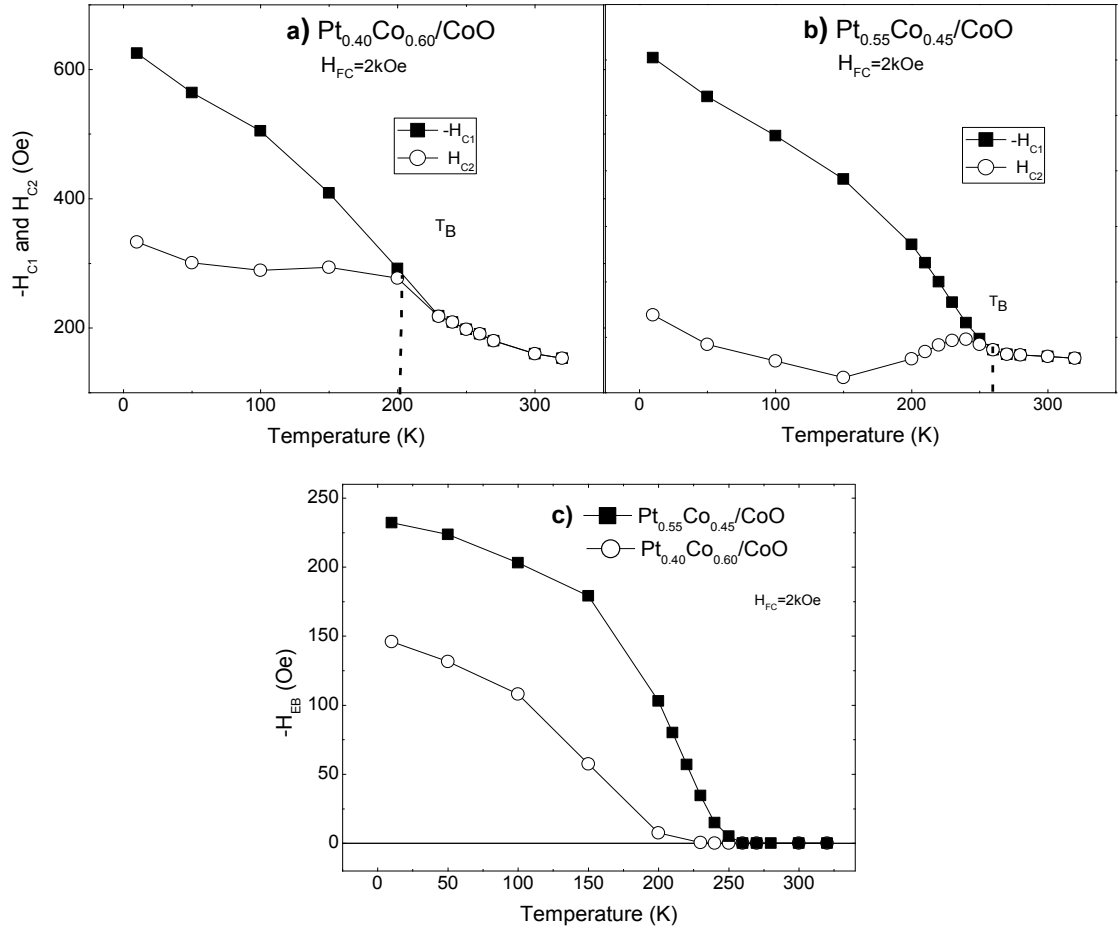

**Fig. 6.** The change of coercive and exchange bias fields as a function of temperature from 10 K to 320 K (Points are the data and lines are guide to the eyes.). a) Temperature dependence of coercive fields for $Pt_{0.40}Co_{0.60}/CoO$. $T_B$ is marked around 200 K from the splitting of $-H_{C1}$ and $H_{C2}$ fields. b) Temperature dependence of coercive fields for $Pt_{0.55}Co_{0.45}/CoO$. $T_B$ is marked around 250 K from the splitting $-H_{C1}$ and $H_{C2}$ fields. c) $-H_{EB}$ fields increase dramatically below $T_B$ for both samples.

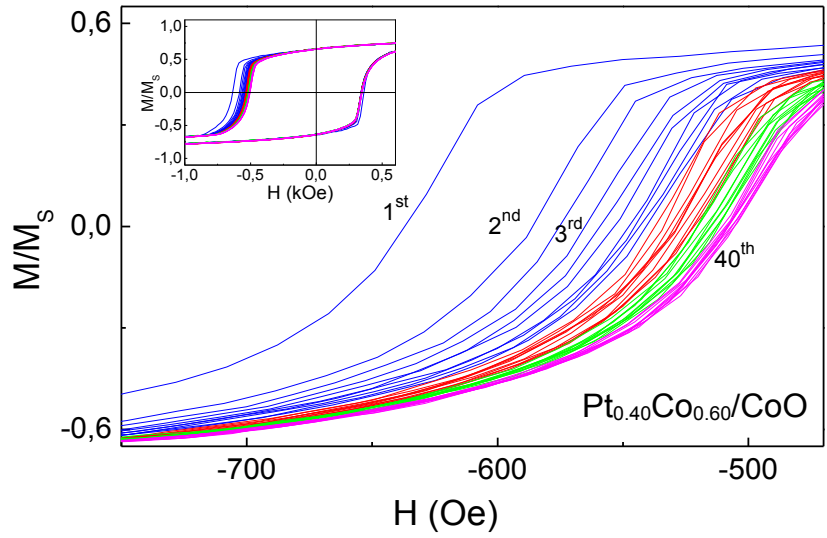

**Fig. 7.** Training effect measurements of $Pt_{0.40}Co_{0.60}/CoO$ sample taken at 10K. The number of consecutive hysteresis loops is 40. The main graph shows the drastic change in $H_{c1}$ and inset shows the full hysteresis loops.

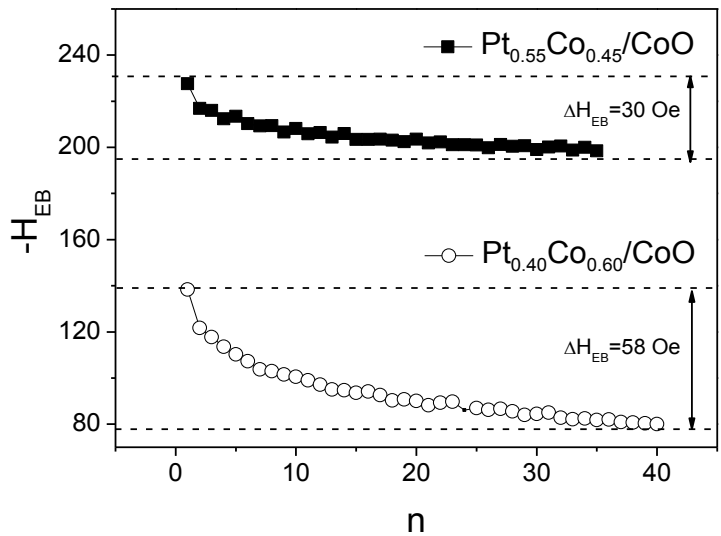

**Fig. 8.** Training of exchange bias ($H_{EB}$) as a function of cycle number (n) for $Pt_{0.40}Co_{0.60}/CoO$ and $Pt_{0.55}Co_{0.45}/CoO$ samples at 10 K.

**Table 1**. Dependence of exchange bias field and blocking temperature on Pt concentration for $Pt_{0.4}Co_{0.6}/CoO$ and $Pt_{0.5}Co_{0.5}/CoO$ samples.

| Pt Concentration | $-H_{EB}$ (Oe) at 10 | $T_B$ (K) |
|:---:|:---:|:---:|
| 0.40 | 140 | 200 |
| 0.55 | 232 | 250 |